\documentclass{jstatphys}
\usepackage[reqno,sumlimits]{amsmath}
\usepackage{amsfonts}
\usepackage{amssymb}
\usepackage{bbm}
\usepackage{graphicx}
\usepackage{psfrag}

\def\R{{\mathbbm{R}}}
\def\RP{{\mathbbm{RP}}}
\def\S{{\mathbb{S}}}
\def\d{{\mathrm{d}}}
\def\e{{\mathrm{e}}}

\begin{document}

\begin{flushright}FAU--TP3--05/2\end{flushright}

\title{On the mean-field spherical model}

\author{Michael Kastner%
\footnote{Physikalisches Institut, Lehrstuhl f\"ur Theoretische Physik I, Universit\"at 
Bayreuth, 95440 Bayreuth, Germany; e-mail: {\tt Michael.Kastner@uni-bayreuth.de}}
and Oliver Schnetz%
\footnote{Institut f\"ur Theoretische Physik III, Friedrich-Alexander-Universit\"at Erlangen-N\"urnberg, Staudtstra{\ss}e 7, 91058 Erlangen, Germany}
}
\runningauthor{Michael Kastner and Oliver Schnetz}

\date{August 22, 2005}

\begin{abstract}
Exact solutions are obtained for the mean-field spherical model, with or without an external magnetic field, for any finite or infinite number $N$ of degrees of freedom, both in the microcanonical and in the canonical ensemble. The canonical result allows for an exact discussion of the {\em loci}\/ of the Fisher zeros of the canonical partition function. The microcanonical entropy is found to be nonanalytic for arbitrary finite $N$. The mean-field spherical model of finite size $N$ is shown to be equivalent to a mixed isovector/isotensor $\sigma$-model on a lattice of two sites. Partial equivalence of statistical ensembles is observed for the mean-field spherical model in the thermodynamic limit. A discussion of the topology of certain state space submanifolds yields insights into the relation of these topological quantities to the thermodynamic behavior of the system in the presence of ensemble nonequivalence.
\end{abstract}


\section{Introduction}

Why discussing in detail an already simple model under even more simplifying assumptions? The spherical model is a simple model, constructed such as to be exactly solvable. In this article, this model is discussed under even more simplifying assumptions, namely mean-field type interactions, coupling each particle to each other at equal strength. The answer to the above question is manifold:
\begin{enumerate}
\item The simplicity of the model allows for an exact solution, with or without external magnetic field, not only in the thermodynamic limit, but also for any finite number $N$ of degrees of freedom, both in the microcanonical (section \ref{sec:mic_sol}) and in the canonical ensemble (section \ref{sec:can_sol}). From these solutions, the calculation of finite-size quantities like the zeros of the canonical partition function is possible (section \ref{sec:Fisher_zeros}).
\item With the exact result for the microcanonical entropy density $s_N$, an example is at hand showing explicitly the nonanalyticity of $s_N$ as a function of the energy {\em for arbitrary finite $N$} (section \ref{sec:nonanalyticity}). This behavior is contrary to the well-known analyticity of the canonical free energy for finite systems. 
\item Although the microcanonical and the canonical ensemble describe different physical situations, both ensembles are known to yield equivalent results in the thermodynamic limit $N\to\infty$ under suitable conditions (like the sufficiently short-rangedness of the interaction potential).\cite{Ruelle} Systems with mean-field type interactions can be viewed as a caricature of (physical) long-range interactions, and they can serve as a playground for the investigation of ensemble nonequivalence.\cite{Dauxois_etal} The mean-field spherical model is found to display a peculiar type of nonequivalence, a phenomenon termed {\em partial equivalence of ensembles} (section \ref{sec:nonequivalence}).
\item Two different statistical mechanical models are considered equivalent when their thermodynamic potentials are equivalent. The known examples of equivalent pairs of models are systems which, beside their differences in the Hamiltonian function, have important aspects in common, like lattice dimensionality and (infinite) system size. In section \ref{sec:mod_equ}, the mean-field spherical model for finite $N$ is shown to be equivalent to a one-dimensional model, the so-called mixed isovector/isotensor $\sigma$-model on two lattice sites. This result implies equivalence of the mean-field spherical model with zero magnetic field to the $\RP^{N-1}$ $\sigma$-model.
\item In a series of papers that appeared in the last years,\cite{CaCaClePe,CaPeCo1,CaCoPe,Angelani_etal1,CaPeCo2,FraPe,GriMo,Angelani_etal2,Kastner2,FraPeSpi,BaCa} topology changes within a family of certain level sets of the state space of a statistical mechanical model were found to be intimately connected to phase transitions. In section \ref{sec:topology}, exact results on topological quantities of the mean-field spherical model are confronted with the thermodynamic results, clarifying the relation between this topological approach to phase transitions and thermodynamics in the presence of ensemble nonequivalence.\footnote{Due to their relevance for the interpretation of the topological approach to phase transitions, part of the results discussed in section \ref{sec:topology} has already been reported (without derivations) in Ref.~\cite{Kastner2}.}
\end{enumerate}
Altogether, its simplicity and the long-rangedness render the mean-field spherical model an instructive and profitable laboratory for the investigation of a variety of effects in the field of the statistical mechanics of phase transitions.

\section{Mean-field spherical model}

The spherical model of a ferromagnet was introduced by T.\ H.\ Berlin and M.\ Kac\cite{BerKac} in 1952. It is devised such as to mimic some features of the Ising model, while, in contrast to the Ising model, having a continuous configuration space. The Hamiltonian function of this model is given by
\begin{equation}
H:\quad\begin{cases}\Lambda &\to\R,\\ \sigma &\mapsto-\dfrac{1}{2}\sum\limits_{i,j=1}^N J_{ij}\,\sigma_i \sigma_j - h\sum\limits_{i=1}^N \sigma_i,\end{cases}
\end{equation}
where a state $\sigma=(\sigma_1,\dotsc,\sigma_N)$ is determined by the values of the $N$ degrees of freedom $\sigma_i\in\R$, $i=1,\dotsc,N$. The coupling matrix $J_{ij}$ determines the strength of the interactions between the $i$-th and the $j$-th degree of freedom, and $h$ is a spatially homogeneous external magnetic field. The state space $\Lambda$ is an ($N$--1)-sphere with radius $\sqrt{N}$, and it is for this reason that the model is called ``spherical''. 

Berlin and Kac considered the case of ferromagnetic nearest-neighbor interactions, where $J_{ij}=J>0$ for all degrees of freedom being nearest neighbors on a $d$-dimensional cubic lattice, and  $J_{ij}=0$ otherwise. They obtained exact expressions for the canonical free energy density $f$ as a function of the temperature $T$. A phase transition takes place for $d\leqslant3$, as $f(T)$ is found to be a nonanalytic function, whereas no transition occurs for $d=1$ and $d=2$ (for a review of results on the spherical model see Ref.\ \cite{Joyce_review}).

Even in the presence of long-range interactions, decaying algebraically with some power $r^{-q}$ ($q>0$) of the distance $r$ on the lattice, an exact analysis of the spherical model is possible in the thermodynamic limit.\cite{Joyce} The existence or absence of a phase transition, as well as the values of the critical exponents in case of a transition, are found to depend on both, the spatial dimension and the value of $q$.

We consider a mean field-like simplification of the spherical model where all degrees of freedom interact with each other at equal strength. The Hamiltonian function of this model can be written as
\begin{equation}\label{Hamiltonian}
H:\quad\begin{cases}\Lambda &\to\R,\\ \sigma &\mapsto-\dfrac{1}{2N}\left(\sum\limits_{i=1}^N \sigma_i\right)^2 - h\sum\limits_{i=1}^N \sigma_i.\end{cases}
\end{equation}
The state space $\Lambda$ is, again, an ($N$--1)-sphere with radius $\sqrt{N}$, and the coupling constants $J_{ij}=\frac{1}{N}$ ($i,j=1,\dotsc,N$) were chosen system-size dependent in order to guarantee the extensivity of the energy.

In the following we will discuss this model within the microcanonical and the canonical ensemble. Note that the terms ``microcanonical'' and ``canonical'' have been used in connection with the spherical model in a different context,\cite{YanWan} referring to the question whether the spherical constraint $\sum\limits_{i=1}^N \sigma_i^2=N$, confining the states $\sigma$ onto the $(N-1)$-sphere $\Lambda$, holds strictly (microcanonically), or only on average (canonically). More appropriately, these two cases should be regarded as different models instead of different statistical ensembles. In the present article, the notions microcanonical and canonical will be used in their meaning from ensemble theory, referring to the physical situations of fixed energy and fixed temperature, respectively.

\section{Microcanonical solution}
\label{sec:mic_sol}

The fundamental quantity for equilibrium statistical mechanics in the microcanonical ensemble is the {\em microcanonical partition function}\/ or {\em density of states}
\begin{equation}\label{Omega_general}
\Omega_N(\varepsilon)=A_N^{-1}\idotsint\limits_\Lambda \d\sigma_1...\d\sigma_N \,\delta[H(\sigma)-N\varepsilon]
\end{equation}
as a function of the energy density $\varepsilon$ and indexed by the number $N$ of degrees of freedom. Here, $\delta$ denotes the Dirac distribution and
\begin{equation}\label{A_N_general}
A_N=\idotsint\limits_\Lambda \d\sigma_1...\d\sigma_N
\end{equation}
is a normalization constant. The $N$-fold integrals extend over the whole state space $\Lambda$. We have
\begin{equation}
\delta\left(\sqrt{\sum\nolimits_{i=1}^N \sigma_i^2}-\sqrt{N}\right)=2\sqrt{N}\,\delta\left(\sum_{i=1}^N \sigma_i^2-N\right)
\end{equation}
and hence, for the mean-field spherical model (\ref{Hamiltonian}), we can write
\begin{multline}\label{omega1}
\Omega_N(\varepsilon)=A_N^{-1}\idotsint\limits_{-\infty}^{+\infty} \d\sigma_1...\d\sigma_N\\
\times \delta\left[\frac{1}{2N}\left(\sum_{i=1}^N \sigma_i\right)^2 + h\sum_{i=1}^N \sigma_i+N\varepsilon\right]\;2\sqrt{N}\,\delta\left[\sum_{i=1}^N \sigma_i^2-N\right],
\end{multline}
where
\begin{equation}\label{A_N}
A_N = \idotsint\limits_{-\infty}^{+\infty} \d\sigma_1...\d\sigma_N \,2\sqrt{N}\delta\left(\sum_{i=1}^N \sigma_i^2-N\right)=\frac{2\pi^{N/2}N^{(N-1)/2}}{\Gamma(N/2)}
\end{equation}
is the surface of an $(N-1)$-sphere of radius $\sqrt{N}$, and $\Gamma$ denotes the (complete) Gamma function. The second Dirac distribution in (\ref{omega1}) is the spherical constraint which restricts the states $\sigma$ to the state space $\Lambda$. By the properties of the Dirac distribution, the integrand in (\ref{omega1}) can be written as
\begin{equation}
\frac{\Theta(h^2-2\varepsilon)}{\sqrt{h^2-2\varepsilon}}\sum_{j=1}^2 \delta\left(\sum_{i=1}^N \sigma_i +Nh_j(\varepsilon)\right)\;2\sqrt{N}\,\delta\left(\sum_{i=1}^N \sigma_i^2-N\right),
\end{equation}
where $\Theta$ denotes the Heaviside step function and
\begin{equation}\label{h12}
h_{1/2}(\varepsilon)=h\pm\sqrt{h^2-2\varepsilon}.
\end{equation}
After interchanging the sum over $j$ and the integral, we shift $\sigma_i\to\sigma_i-h_j(\varepsilon)$ and obtain
\begin{multline}\label{omega2}
\Omega_N(\varepsilon)=\frac{\Theta(h^2-2\varepsilon)}{A_N\sqrt{h^2-2\varepsilon}} \sum_{j=1}^2  \idotsint\limits_{-\infty}^{+\infty} \d\sigma_1...\d\sigma_{N-1}\\
\times2\sqrt{N}\,\delta\left[\left(\sum_{i=1}^{N-1}\sigma_i\right)^2+\sum_{i=1}^{N-1}\sigma_i^2-N\left(1-h_j^2(\varepsilon)\right)\right]
\end{multline}
by integrating over $\sigma_N$. The argument of the Dirac distribution contains an $(N-1)\times(N-1)$ bilinear form,
\begin{equation}
\left(\sum_{i=1}^{N-1}\sigma_i\right)^2+\sum_{i=1}^{N-1}\sigma_i^2 =
\begin{pmatrix}\sigma_1\\ \vdots\\ \sigma_{N-1}\end{pmatrix}^{\!\!\!\sf T}{\mathcal A}\begin{pmatrix}\sigma_1\\ \vdots\\ \sigma_{N-1}\end{pmatrix},
\end{equation}
where ${\mathcal A}$ is an $(N-1)\times(N-1)$ matrix with entries ${\mathcal A}_{k\ell}=1+\delta_{k,\ell}$, and $\delta$ denotes Kronecker's delta. The eigenvalues of ${\mathcal A}$ are $N$ and $1$, where the latter is $(N-2)$-fold degenerate. Hence, by an orthogonal transformation of variables in (\ref{omega2}) and a rescaling in the direction of the eigenvalue $N$, we obtain
\begin{multline}\label{omega3}
\Omega_N(\varepsilon)=\frac{\Theta(h^2-2\varepsilon)}{A_N\sqrt{N(h^2-2\varepsilon)}}\sum_{j=1}^2 \idotsint\limits_{-\infty}^{+\infty} \d\sigma_1...\d\sigma_{N-1}\\
\times2\sqrt{N}\,\delta\left[\sum_{i=1}^{N-1}\sigma_i^2-N\left(1-h_j^2(\varepsilon)\right)\right].
\end{multline}
For $h_j^2(\varepsilon)<1$, the integral in (\ref{omega3}) describes the surface of an $(N-2)$-sphere of radius $\sqrt{N\left(1-h_j^2(\varepsilon)\right)}$ (and vanishes otherwise). With equation (\ref{A_N}) we obtain
\begin{equation}\label{omega_final}
\Omega_N(\varepsilon)=\frac{\Gamma(\frac{N}{2})}{\Gamma(\frac{N-1}{2})}\, \frac{\Theta(h^2-2\varepsilon)}{N\sqrt{\pi(h^2-2\varepsilon)}} \sum_{j=1}^2 \left[1-h_j^2(\varepsilon)\right]^{\frac{N-3}{2}}\Theta[1-h_j^2(\varepsilon)]
\end{equation}
as a final expression for the microcanonical partition function of the mean-field spherical model (\ref{Hamiltonian}), where $h_j$ is defined in (\ref{h12}).

\begin{figure}[tb]
\begin{center}
\psfrag{e}{$\varepsilon$}
\psfrag{h}{$h$}
\psfrag{-}{$-$}
\psfrag{1}{1}
\psfrag{2}{2}
\includegraphics[width=7.2cm,height=6.6cm,clip=false]{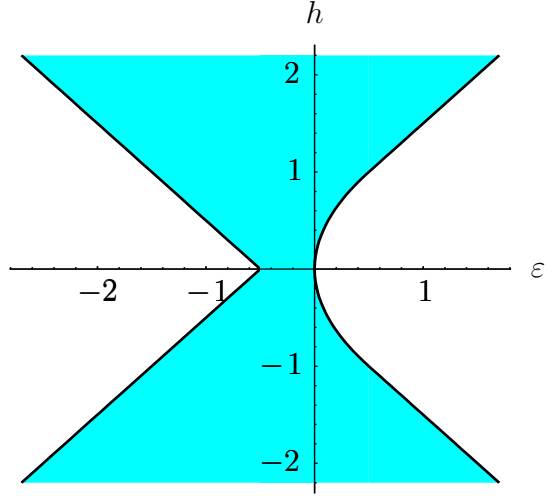}
\caption{\label{fig:domain} \small
Domain of the microcanonical entropy density $s(\varepsilon)$ for the mean-field spherical model (\ref{Hamiltonian}) in dependence of the value of the external magnetic field $h$. Points within the shaded region obey $-1<|h|-\sqrt{h^2-2\varepsilon}<1$ and correspond to accessible states.
}
\end{center}
\end{figure}

\begin{figure}[tb]
\psfrag{e}{\small $\varepsilon$}
\psfrag{s}{\small $s$}
\psfrag{h}{\small $h$}
\psfrag{=1}{\small $=\!1$}
\psfrag{=0}{\small $=\!0$}
\psfrag{-1.5}[c][][0.8]{\small \!\!\!\!\!\!$-1.5$}
\psfrag{-1.0}[c][][0.8]{\small \!\!\!\!\!\!\!\!\!\!\!\!$-1.0$}
\psfrag{-0.5}[c][][0.8]{\small \!\!\!\!\!\!$-0.5$}
\psfrag{0.5}[c][][0.8]{\small \!\!\!\!$0.5$}
\psfrag{-0.4}[c][][0.8]{\small \!\!\!\!\!\!$-0.4$}
\psfrag{-0.3}[c][][0.8]{\small \!\!\!\!\!\!$-0.3$}
\psfrag{-0.2}[c][][0.8]{\small \!\!\!\!\!\!$-0.2$}
\psfrag{-0.1}[c][][0.8]{\small \!\!\!\!\!\!$-0.1$}
\psfrag{-1}[c][][0.8]{\small $-1$}
\psfrag{-2}[c][][0.8]{\small $-2$}
\psfrag{-3}[c][][0.8]{\small $-3$}
\begin{center}
\includegraphics[width=5.5cm,height=4cm,clip=false,keepaspectratio=true]{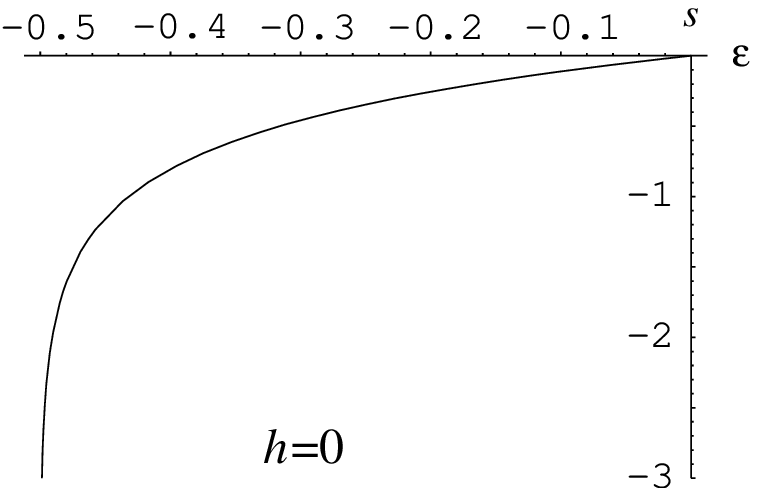}
\hspace{6mm}
\includegraphics[width=5.5cm,height=4cm,clip=false,keepaspectratio=true]{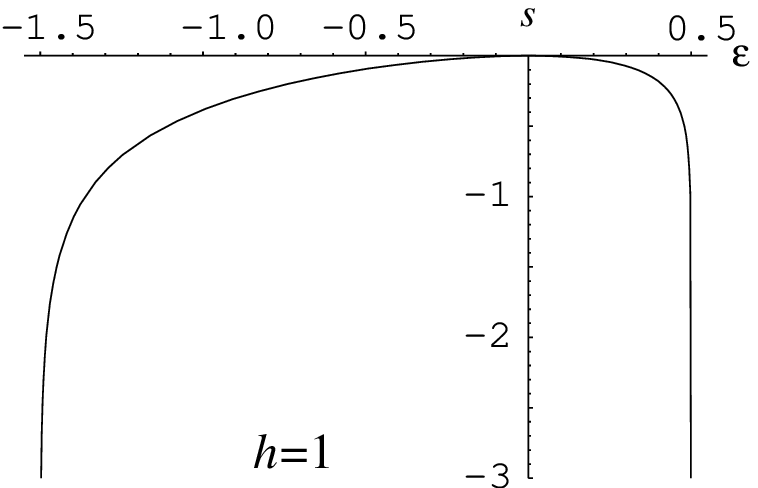}
\caption{\label{fig:svone} \small
Exemplary plots of the graph of the microcanonical entropy density $s$ for the cases $h=0$ and $h=1$. The domain of $s$ is the interval $(-\frac{1}{2},0]$ for $h=0$, and $(-\frac{3}{2},\frac{1}{2})$ for $h=1$. Note that, for $h=0$, the slope $\frac{\partial s(\varepsilon)}{\partial \varepsilon}\geqslant 1$ has a positive lower bound, whereas the slope is unbounded for all $h\neq0$.
}
\end{center}
\end{figure}

The microcanonical entropy density is defined as
\begin{equation}
s_N(\varepsilon)=\frac{1}{N}\ln\Omega_N(\varepsilon).
\end{equation}
Its thermodynamic limit,
\begin{equation}
s(\varepsilon)=\lim_{N\to\infty}s_N(\varepsilon),
\end{equation}
is the microcanonical counterpart of canonical thermodynamic potentials like the Gibbs free energy density, and it constitutes the starting point for microcanonical thermodynamics. Making use of Sterling's formula, we obtain from the microcanonical partition function (\ref{omega_final})
\begin{equation}\label{svone}
s(\varepsilon)=\frac{1}{2}\ln\left[1-\left(|h|-\sqrt{h^2-2\varepsilon}\right)^2\right],
\end{equation}
defined for all values of $\varepsilon$ and $h$ for which the radical and the argument of the logarithm are positive. For visualization, in figure \ref{fig:domain} the domain of $s$ is illustrated in the $(\varepsilon,h)$-plane. 
Exemplary plots of the graph of $s$ with and without external magnetic field $h$, respectively, are shown in figure \ref{fig:svone}.

\section{Canonical solution}
\label{sec:can_sol}

The fundamental quantity for equilibrium statistical mechanics in the canonical ensemble is the {\em canonical partition function}
\begin{equation}\label{Z_general}
Z_N(\beta)=A_N^{-1}\idotsint\limits_\Lambda \d\sigma_1...\d\sigma_N \,\e^{-\beta H(\sigma)}
=N\int\limits_{-\infty}^{+\infty} \d \varepsilon\, \Omega_N(\varepsilon)\, \e^{-\beta N\varepsilon}
\end{equation}
as a function of the inverse temperature $\beta=\frac{1}{k_B T}>0$, where $k_B$ is Boltzmann's constant and $T$ is the temperature. The normalization constant $A_N$ is the same as in (\ref{A_N_general}), and the integration, as in the definition of the microcanonical partition function (\ref{Omega_general}), extends over the whole state space $\Lambda$. Inserting the microcanonical partition function (\ref{omega_final}) into (\ref{Z_general}) and interchanging the sum over $j$ with the integral, the canonical partition function reads
\begin{equation}\label{Z_N}
Z_N(\beta)=\frac{\Gamma(\frac{N}{2})}{\Gamma(\frac{N-1}{2})\sqrt{\pi}}\int\limits_{-1}^1 \d z\,\left(1-z^2\right)^{\frac{N-3}{2}}\exp\left\{\beta Nz\left(\tfrac{z}{2}+ h\right)\right\}
\end{equation}
where $z=h_j(\varepsilon)$ has been substituted. This expression can be written as an infinite sum of confluent hypergeometric functions $_1F_1$. For zero external magnetic field $h=0$, a single summand survives, yielding
\begin{equation}\label{Z_N_h=0}
Z_N^{h=0}(\beta)=\,\!_1F_1\!\left({\textstyle \frac{1}{2},\frac{N}{2},\frac{\beta N}{2}}\right)
\end{equation}
for the zero-field canonical partition function of the mean-field spherical model.

The canonical free energy density is defined as
\begin{equation}\label{f_N}
f_N(\beta)=-\frac{1}{N\beta}\ln Z_N(\beta).
\end{equation}
Its thermodynamic limit,
\begin{equation}
f(\beta)=\lim_{N\to\infty} f_N(\beta),
\end{equation}
constitutes the starting point for canonical thermodynamics. An exact expression for $f(\beta)$ can be obtained in various ways: either by a large $N$-expansion \cite{SoSta} of the integral in (\ref{Z_N_h=0}), or, analogously to the calculations of Berlin and Kac,\cite{BerKac} by making use of the method of steepest descent. Having already at hand an expression for the microcanonical entropy density $s$, the free energy density is most easily obtained by means of a Legendre-Fenchel transform,
\begin{equation}\label{Legendre-Fenchel}
f(\beta)=\inf_\varepsilon \left[\varepsilon-\beta^{-1}s(\varepsilon)\right].
\end{equation}
Evaluating this expression, we find that
\begin{equation}
f(\beta)=\bar{\varepsilon}(\beta) - \beta^{-1}s[\bar{\varepsilon}(\beta)]
\end{equation}
with the following expressions for $\bar{\varepsilon}(\beta)$:

For the case of zero external magnetic field $h=0$, 
\begin{equation}\label{evonbeta_h=0}
\bar{\varepsilon}(\beta)=\begin{cases}
0 & \text{for $\beta\leqslant1$,}\\
\frac{1-\beta}{2\beta} & \text{for $\beta>1$,}
\end{cases}
\end{equation}
yielding
\begin{equation}\label{fvonbeta_h=0}
f(\beta)=\begin{cases}
0 & \text{for $\beta\leqslant1$,}\\
\frac{1-\beta+\ln\beta}{2\beta} & \text{for $\beta>1$,}
\end{cases}
\end{equation}
for the canonical free energy density for zero external magnetic field. The cusps in the graphs of $\bar{\varepsilon}(\beta)$ and $f(\beta)$ can be viewed as a consequence of the fact that, for $h=0$, the microcanonical entropy density (\ref{svone}) has a compact support and its slope has a positive lower bound (see figure \ref{fig:svone}).

\begin{figure}[tb]
\begin{center}
\psfrag{b}{\small $\beta$}
\psfrag{f}{\small $f$}
\psfrag{h=0}{\small $h\!=\!0$}
\psfrag{h=1}{\small $h\!=\!1$}
\psfrag{-1.2}[c][][0.8]{\small \!\!$-1.2$}
\psfrag{-1.0}[c][][0.8]{\small \!\!$-1.0$}
\psfrag{-0.8}[c][][0.8]{\small \!\!$-0.8$}
\psfrag{-0.6}[c][][0.8]{\small \!\!$-0.6$}
\psfrag{-0.4}[c][][0.8]{\small \!\!$-0.4$}
\psfrag{-0.3}[c][][0.8]{\small \!\!$-0.3$}
\psfrag{-0.2}[c][][0.8]{\small \!\!$-0.2$}
\psfrag{-0.1}[c][][0.8]{\small \!\!$-0.1$}
\psfrag{2}[c][][0.8]{\small $2$}
\psfrag{4}[c][][0.8]{\small $4$}
\psfrag{6}[c][][0.8]{\small $6$}
\psfrag{8}[c][][0.8]{\small $8$}
\psfrag{10}[c][][0.8]{\small $10$}
\includegraphics[width=5.5cm,height=4cm,clip=false,keepaspectratio=true]{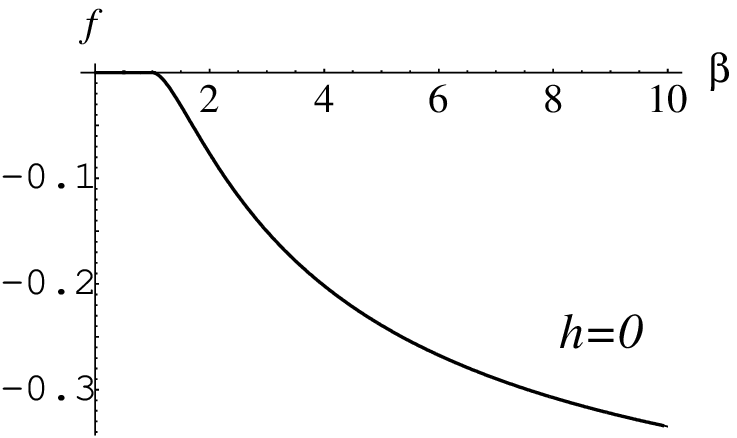}
\hspace{6mm}
\includegraphics[width=5.5cm,height=4cm,clip=false,keepaspectratio=true]{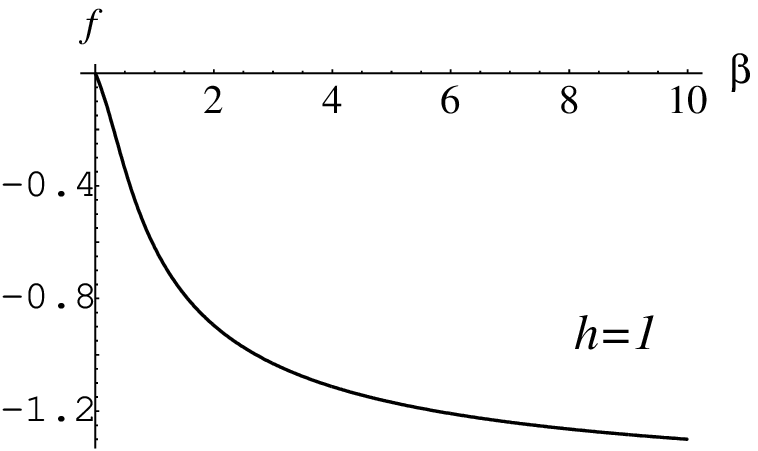}
\caption{\label{fig:fvonbeta} \small
The graph of the canonical free energy density $f$, showing a nonanalyticity at $\beta=1$ for $h=0$, whereas $f$ is smooth for $h\neq0$. 
}
\end{center}
\end{figure}

For nonzero field $h$, $\bar{\varepsilon}(\beta)$ can be expressed as the real root of a third order polynomial in $\bar{\varepsilon}$,
\begin{multline}
8\beta^2 \bar{\varepsilon}^3-4\beta(\beta h^2-2\beta+2)\bar{\varepsilon}^2-2(6\beta^2 h^2-\beta^2-2\beta h^2+2\beta-1)\bar{\varepsilon}\\+\beta h^2(4\beta h^2-\beta+2)=0.
\end{multline}
In both cases, the expressions for $\bar{\varepsilon}(\beta)$ are derived from the inverse function of the positive part [$\beta(\varepsilon)>0$] of the microcanonical inverse temperature
\[
\beta(\varepsilon)=\frac{\partial s(\varepsilon)}{\partial \varepsilon}=\frac{1+2\varepsilon-(1-2\varepsilon)h(h^2-2\varepsilon)^{-1/2}}{(1+2\varepsilon)^2-4h^2}.
\]
Exemplary plots of the graph of $f$ are shown in figure \ref{fig:fvonbeta}.

Physically, the observed nonanalyticity in the zero field case marks a transition from a ferromagnetic phase with nonvanishing magnetization at low temperatures ($\beta>1$) towards a paramagnetic phase with zero magnetization at high temperatures ($\beta<1$).

\section{Equivalence to an isovector/isotensor $\boldsymbol{\sigma}$-model}
\label{sec:mod_equ}

Two different models are considered equivalent when they have identical thermodynamic potentials. Well-known examples of systems being equivalent in the thermodynamic limit include
\begin{enumerate}
\item a lattice gas of (quantum) particles, whose grand canonical potential equals the canonical thermodynamic potential of an (anisotropic quantum) spin-$\frac{1}{2}$ system in an external magnetic field on the same lattice,\cite{LeeYang2,MatMat}
\item the $D$-vector model (comprising the Ising, the $XY$-, and the Heisenberg model as cases $D=1,2,3$, respectively) which, on a hypercubic lattice and in the limit $D\to\infty$, is equivalent to the spherical model on the same lattice.\cite{Stanley,KacThom}
\end{enumerate}
Note, however, that in general such equivalence does not extend to other quantities apart from thermodynamic potentials and quantities derived therefrom.\cite{WeiJa}

A more intricate relation is found for the mean-field spherical model and an extension of the $\RP^{D-1}$ $\sigma$-model, the so-called mixed isovector/isotensor $\sigma$-model. The $\RP^{D-1}$ $\sigma$-model is a lattice model with an order parameter taking values in the real projective space $\RP^{D-1}$. This system has attracted some interest as a model of nematic liquid crystals,\cite{KoShro} as well as in the particle physics community as a laboratory for investigating asymptotic freedom, topological phenomena, and universality. For a list of references and an account of the history of the $\RP^{D-1}$ $\sigma$-model, the reader is invited to consult the article by Sokal and Starinets.\cite{SoSta} In the same reference, the following extension of this model is discussed.

The mixed isovector/isotensor $\sigma$-model is defined by the lattice Hamiltonian
\begin{equation}\label{H_sigma}
H:\quad\begin{cases} \left(\S^{D-1}\right)^N & \to\R, \\ \sigma & \mapsto -D\sum\limits_{\langle i,j\rangle}\left[h\,\sigma_i \cdot \sigma_j + \frac{1}{2}\left(\sigma_i \cdot \sigma_j\right)^2\right],\end{cases}
\end{equation}
where the sum runs over all pairs $\langle i,j\rangle$ (each pair counted once) of nearest neighbors on a lattice. Each component $\sigma_i$ of a state vector $\sigma=(\sigma_1,\dotsc,\sigma_N)$ represents a point on a $(D-1)$-dimensional unit sphere $\S^{D-1}$. For $h=0$, (\ref{H_sigma}) reduces to the Hamiltonian of the $\RP^{D-1}$ $\sigma$-model. In reference \cite{SoSta}, the canonical free energy density of the mixed isovector/isotensor $\sigma$-model for arbitrary $D$ on a one-dimensional lattice of size $N$ with open boundary conditions was shown to be of the form
\begin{equation}
Z_{N,D}(\beta)=\left[\frac{\Gamma(\frac{D}{2})}{\Gamma(\frac{D-1}{2})\sqrt{\pi}}\int\limits_{-1}^1 \d z\,\left(1-z^2\right)^{\frac{D-3}{2}}\exp\left\{\beta Dz\left(h+\frac{z}{2}\right)\right\}\right]^{N-1}.
\end{equation}
Upon setting $N=2$ in this expression and identifying $D$ with $N$, the canonical partition function (\ref{Z_N}) of the mean-field spherical model is recovered and the two models under consideration are found to be equivalent. Remarkably, and unlike the cases of model equivalence mentioned above, the models differ in fundamental aspects: The mean-field spherical model of finite size $N$ with an external magnetic field is equivalent to the mixed isovector/isotensor $\sigma$-model with $D=N$ on a (one-dimensional) lattice of only two sites. For the sake of clarity, the characteristics of the two equivalent models are confronted in the following table:
\begin{center}
\begin{tabular}{|l|c|c|}
\hline
model: & mean-field spherical & $\RP^{N-1^{\vphantom{1}}}$ $\sigma$\\
\hline
lattice dimension $d$: & --- & 1\\
system size: & $N$ & 2\\
degrees of freedom per site: & 1 & $N-1$\\
\hline
\end{tabular}
\end{center}
Note that, for the two models considered above, model equivalence can be established even on the level of the Hamiltonian functions. The Hamiltonian (\ref{H_sigma}) of the mixed isovector/isotensor $\sigma$-model consisting of two lattice sites can be mapped onto the Hamiltonian (\ref{Hamiltonian}) of the mean-field spherical model by first rescaling $\sigma_1$ and $\sigma_2$ by a factor of $D^{-1/2}$, and then exploiting the rotational symmetry of the model by setting $\sigma_1=(1,\dotsc,1)$.

\section{Fisher zeros of the canonical partition function}
\label{sec:Fisher_zeros}

The rare situation that a statistical system of $N$ interacting degrees of freedom can be solved for arbitrary $N$ allows to discuss the exact location of the so-called Fisher zeros,\cite{Fisher} the zeros of the canonical partition function in the complex temperature plane. The Fisher zeros form the natural analogon to the zeros of the grand canonical partition function in the complex fugacity plane as introduced by Yang and Lee in their seminal article.\cite{LeeYang1}

\begin{figure}[tb]
\begin{center}
\psfrag{1.1}{1.1}
\psfrag{1.2}{1.2}
\psfrag{1.3}{1.3}
\psfrag{0.2}{0.2}
\psfrag{0.4}{0.4}
\psfrag{-0.2}{\!\!$-0.2$}
\psfrag{-0.4}{\!\!$-0.4$}
\psfrag{Reb}{$\Re(\beta)$}
\psfrag{Imb}{$\Im(\beta)$}
\includegraphics[width=9.5cm,height=8.5cm,clip=false,keepaspectratio=true]{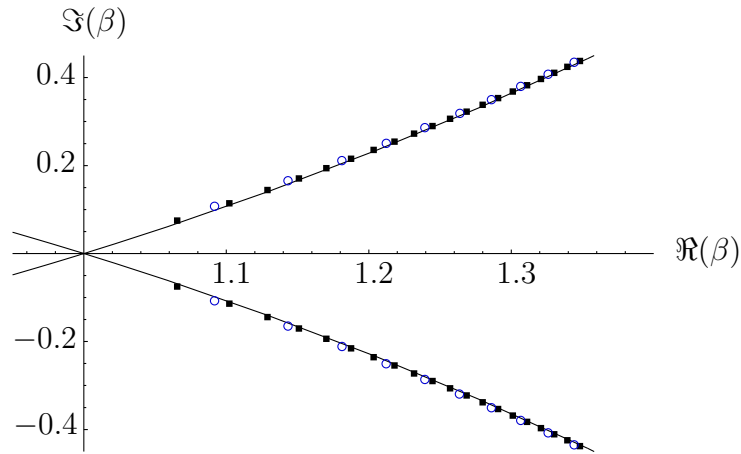}
\caption{\label{fig:complexzeros} \small
Szeg\"o curve (\ref{szego}) in the complex $\beta$-plane, together with some zeros of $_1F_1(\frac{1}{2},\frac{N}{2},\frac{N\beta}{2})$ for $N=1000$ (circles) and $N=2000$ (squares).
}
\end{center}
\end{figure}

For the mean-field spherical model, a discussion of the Fisher zeros of the zero-field canonical partition function (\ref{Z_N_h=0}) amounts to analyzing the complex zeros of the confluent hypergeometric function $_1F_1(\frac{1}{2},\frac{N}{2},\cdot)$ with respect to its third argument. Focusing on the large $N$ asymptotics, we are interested in the complex zeros of $_1F_1(\frac{1}{2},\frac{N}{2},\frac{N\beta}{2})$ with respect to $\beta$ in the limit of large $N$. At this point, we can benefit from the work by Sokal and Starinets:\cite{SoSta} They proved that, as illustrated in figure \ref{fig:complexzeros}, the complex zeros lie close to the two branches with real part $\Re(\beta)>1$ of the Szeg\"o curve
\begin{equation}\label{szego}
|\beta|^2=\e^{2(\Re(\beta)-1)},
\end{equation}
approaching the curve and the value $\beta=1$ in the thermodynamic limit of large $N$.
This is a beautiful illustration of the Yang-Lee-mechanism of analyticity breaking, in which, for large but finite $N$, the complex zeros of a real-analytic function (the canonical partition function) approach and pinch the real axis, giving rise to a nonanalyticity in the limit $N\to\infty$. More precisely, in this limit the zeros lie dense on the branches of the Szeg\"o curve, thus cutting the complex plane into two disconnected domains.

\section{Nonanalyticity of the finite-system entropy}
\label{sec:nonanalyticity}

As mentioned in the preceding section, the canonical partition function, and hence also the canonical free energy density $f_N$ as defined in (\ref{f_N}), are known to be real-analytic functions of $\beta$ for any finite number $N$ of degrees of freedom. For microcanonical thermodynamic functions like the entropy density $s_N$, an argument guaranteeing analyticity does not exist. Nonetheless, the belief that $s_N$ were real-analytic in $\varepsilon$, serving as a justification for certain arguments or operations, has been expressed in several publications.\cite{Behringer,BePleiHue}

With the result (\ref{omega_final}) for the microcanonical partition function $\Omega_N$ of the mean-field spherical model, we have an exact result at hand from which it is straightforward to read off the nonanalyticity of $\Omega_N$, and therefore of $s_N$, for all values of the external magnetic field $h$ satisfying $0<|h|<1$. The nonanalyticity is located at $\varepsilon=|h|-\frac{1}{2}$, as at this value the number of nonvanishing contributions in the sum over $j$ in (\ref{omega_final}) jumps from one to two. The nonanalyticity is such that a discontinuity occurs in the $\lambda$-th derivative of $s_N(\varepsilon)$ with respect to $\varepsilon$, where
\begin{equation}
\lambda=\lambda(N)=\left\lfloor\frac{N-2}{2}\right\rfloor,
\end{equation}
i.\,e., the largest integer smaller than or equal to $\frac{N-2}{2}$. The observation that $\lambda$ is of order $N$ for $N\to\infty$ is in agreement with (\ref{svone}) where we found a smooth microcanonical entropy density $s$ in the thermodynamic limit.

One might argue that such a nonanalytic behavior of $s$ is of little physical relevance for the system sizes $N$ under consideration in statistical physics. For suitable choices of the interaction potential, however, we expect to find models of physical interest for which $\lambda$ is of order 1.\cite{Kastner_wip}

\section{Nonequivalence of statistical ensembles}
\label{sec:nonequivalence}

As already pointed out in the introduction, the sufficiently short-rangedness of the interaction potential is a sufficient condition for the equivalence of the statistical ensembles in the thermodynamic limit. If, however, the microcanonical entropy density $s$ is explicitely known, more can be said: the equivalence, or nonequivalence, of statistical ensembles can, loosely speaking, be traced back to the concavity properties of $s$ (see the article by Ellis, Haven, and Turkington\cite{ElHaTur} for a precise formulation of this statement as well as of the definition of ensemble equivalence).

For a third type of relation of the ensembles, situated somewhere in-between equivalence and nonequivalence, the term ``partial equivalence'' has been coined,\cite{ElHaTur} and it is this kind of situation we encounter for the mean-field spherical model with zero external magnetic field $h=0$: The microcanonical entropy density, as depicted in figure \ref{fig:svone}, is a strictly concave function with compact support. Furthermore, we find $\beta(\varepsilon)=\frac{\partial s(\varepsilon)}{\partial \varepsilon}\in[1,\infty)$, i.\,e., a positive lower bound exists on the microcanonical inverse temperature. As a consequence of these properties and of the fact that the canonical free energy density $f$ can be derived from $s$ be means of a Legendre-Fenchel-transform (\ref{Legendre-Fenchel}), the following relation arises: For all values of $\beta\geqslant 1$ above the lower bound on the microcanonical inverse temperature, a one-to-one correspondence between microcanonical and canonical macrostates exists, and the two ensembles are said to be equivalent. For $\beta<1$, however, all the various canonical macrostates correspond to the microcanonical macrostate with maximum energy $\varepsilon=0$. This is clearly not a one-to-one correspondence and the ensembles are nonequivalent for this range of inverse temperatures. It is in this sense that, in spite of the strict concavity of the microcanonical entropy density $s$, the equivalence between the ensembles is only ``partial'' for the mean-field spherical model.

For arbitrary nonzero values of the external magnetic field $h\neq0$, the microcanonical entropy density is strictly concave and $\frac{\partial s(\varepsilon)}{\partial \varepsilon}$ is unbounded below and above, hence equivalence of the microcanonical and the canonical ensemble holds.

Focusing on the question whether a phase transition is present in one ensemble or the other, the following considerations are of interest: one of the common choices is to define a phase transition as the presence of a nonanalyticity in the canonical free energy density. For all finite $N$, this quantity is known to be a real-analytic function of the (inverse) temperature. Only in the thermodynamic limit $N\to\infty$ a nonanalyticity, and hence a phase transition, may be found. This is exactly the situation we observe for the mean-field spherical model with zero external magnetic field $h=0$: For systems consisting of a finite number $N$ of degrees of freedom, the canonical free energy density $f_N(\beta)$ is the logarithm of a confluent hypergeometric function with all positive arguments [equations (\ref{Z_N_h=0}) and (\ref{f_N})], and is therefore real-analytic. In the thermodynamic limit, however, the nonanalyticity in $f(\beta)$ at $\beta=1$ is obvious from (\ref{fvonbeta_h=0}).

In the microcanonical ensemble, conditions on the microcanonical entropy density $s$ for the existence of a phase transition are somewhat less established. An obvious way to start is to look for conditions on $s$ which, in the case of ensemble equivalence, correspond to the definition of a phase transition from the canonical free energy density.\footnote{A different definition of phase transitions, based on the microcanonical entropy, is proposed in Ref.\ \cite{BouBa}.} Such conditions on $s(\varepsilon)$ are
\begin{enumerate}
\item the existence of an open interval of energies $\varepsilon$ for which $s(\varepsilon)$ is not strictly concave, or
\item a nonanalyticity in $s(\varepsilon)$.
\end{enumerate}
Then it might appear reasonable to adopt the same conditions also for systems for which equivalence of ensembles does not hold. Interpreting the microcanonical entropy density $s$ obtained for $h=0$ in the thermodynamic limit in (\ref{svone}) along the lines of (i) and (ii), it is straightforward to discuss the issue of existence of a phase transition: $s$ is a real-analytic and strictly concave function on its entire domain, therefore no phase transition takes place in the microcanonical ensemble.

The situation observed for the mean-field spherical model with $h=0$ is typical for systems showing partial equivalence of ensembles. (For further examples see reference \cite{BaBoDaRu}.) In spite of the fact that partial equivalence appears to be a relatively ``mild'' form of ensemble nonequivalence, its impact on the existence of phase transitions can be substantial, resulting in the {\em existence}\/ of a phase transition in the canonical ensemble, and the {\em absence}\/ in the microcanonical one.

\section{Topology of state space submanifolds}
\label{sec:topology}

An entirely different approach to phase transitions, not making use of any of the Gibbs ensembles, has been proposed a couple of years ago. This {\em topological approach}\/ connects the occurrence of a phase transition to certain properties of the Hamiltonian function $H$, resorting to topological concepts. From a conceptual point of view, this approach has a remarkable property: The microscopic Hamiltonian dynamics can be linked via the maximum Lyapunov exponent to the topological quantities considered.\cite{CaPeCo1} With the topological approach, in turn, linking a change of the topology to the occurrence of a phase transition, a concept is established which provides a connection between a phase transition in a system and its underlying microscopic dynamics.

The topological approach is based on the hypothesis\cite{CaCaClePe} that phase transitions are related to topology changes of submanifolds $\Sigma_\varepsilon$ of the state space of the system. The $\Sigma_\varepsilon$ consist of all points $q$ of the state space $\Lambda$ for which the energy $H(\sigma)$ per degree of freedom equals a certain level $\varepsilon$, i.\,e.,
\begin{equation}\label{Sigma_def}
\Sigma_\varepsilon=\left\{ \sigma\in\Lambda\,\Big|\,\frac{H(\sigma)}{N}= \varepsilon\right\}.
\end{equation}
(Or, in a related version, the topology of submanifolds $M_\varepsilon$ consisting of all points $q$ with $H(q)/N\leqslant \varepsilon$ is considered.) The hypothesis then conjectures that a topology change within the family $\left\{\Sigma_\varepsilon\right\}$ at $\varepsilon=\varepsilon_c$ is a necessary condition for a thermodynamic phase transition to take place at $\varepsilon_c$ [or at the corresponding critical temperature $T_c=T(\varepsilon_c)$]. This hypothesis has been corroborated by numerical and by exact results for a model showing a first-order phase transition\cite{Angelani_etal1,Angelani_etal2} as well as for systems with second-order phase transitions.\cite{CaPeCo1,GriMo,RiTeiSta,Kastner} A major achievement in the field is the recent proof of a theorem, stating, loosely speaking, that, for systems described by smooth, finite-range, and confining potentials, a topology change of the submanifolds $\Sigma_\varepsilon$ is a {\em necessary}\/ criterion for a phase transition.\cite{FraPe,FraPeSpi} Recent results on the state space topology of the mean-field $\phi^4$-model have cast some doubt on the general relevance of topology changes for the occurrence of a phase transition,\cite{GaSchiSca,AnAnRuoZam} but these findings require further investigation.

A topology change in $\Sigma_\varepsilon$ is clearly not {\em sufficient}\/ to entail a phase transition. This follows for example from the analytical computation of topological invariants in the $k$-trigonometric model,\cite{Angelani_etal1,Angelani_etal2} where the number of topology changes occurring is shown to diverge in the limit $N\to\infty$, although only a single phase transition takes place. So topology changes appear to be rather common, and only particular ones are related to phase transitions. While some ideas relating the ``strength'' of a topology change to the occurrence of a phase transitions have been put forward,\cite{CaPeCo2} there are strong indications that such a criterion based exclusively on topological quantities cannot exist in general.\cite{Kastner}

\begin{figure}[tb]
\begin{center}
\psfrag{v}{$\varepsilon$}
\psfrag{h}{$h$}
\psfrag{-}{$-$}
\psfrag{1}{1}
\psfrag{2}{2}
\includegraphics[width=7.2cm,height=6.6cm,clip=false]{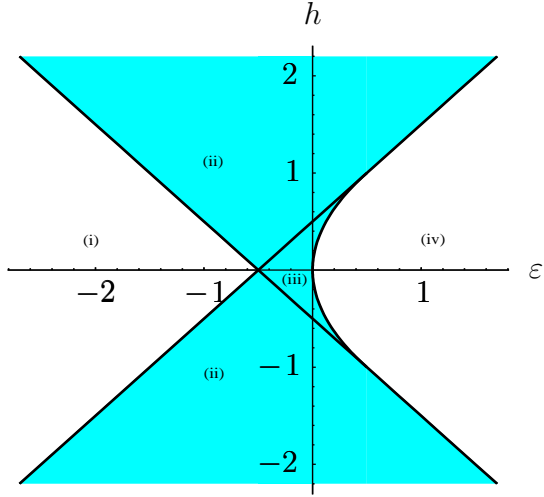}
\caption{\label{changes} \small
Lines in the $(\varepsilon,h)$-plane where topology changes occur. The shaded region marks the support of the density of states $\Omega_N(\varepsilon)$, i.e., the values of $\varepsilon$ which, for a given external field $h$, are accessible for the system. The various regions in the plane separated by the bold lines correspond to the following submanifold topologies: (i) to the left of the shaded region: $\Sigma_\varepsilon\sim\varnothing$, (ii) the two shaded triangles: $\Sigma_\varepsilon\sim\S^{N-2}$, (iii) the small, triangle shaped shaded region:  $\Sigma_\varepsilon\sim\S^{N-2}+\S^{N-2}$, (iv) to the right of the shaded region: $\Sigma_\varepsilon\sim\varnothing$.}
\end{center}
\end{figure}

Ribeiro Teixeira and Stariolo have analyzed the topological structure of the state space submanifolds of the mean-field spherical model in reference \cite{RiTeiSta}. Completing their findings and translating them into our notation, the topology of the submanifolds $\Sigma_\varepsilon$ can be written as
\begin{equation}\label{Sigma_topo}
\Sigma_\varepsilon\sim
\begin{cases}
\S^{N-2} & \text{for $-|h|-\frac{1}{2}<\varepsilon<|h|-\frac{1}{2}$},\\
\S^{N-2}+\S^{N-2} & \text{for $|h|-\frac{1}{2}<\varepsilon<\frac{h^2}{2}$ and $|h|<1$},\\
\varnothing & \text{otherwise},
\end{cases}
\end{equation}
where $\sim$ indicates topological equivalence, the $+$-sign denotes a topological (unconnected) sum, and $\varnothing$ is the empty set. The regions of the $(\varepsilon,h)$-plane corresponding to the different topologies and the lines at which topology changes occur are plotted in figure \ref{changes}.
The only topology change in the interior of the physical parameter domain is from $\S^{N-2}$ to $\S^{N-2}+\S^{N-2}$. This change occurs for $0<|h|<1$ and $\varepsilon=|h|-\frac{1}{2}$ and is responsible for the nonanalyticity of $s_N(\varepsilon)$ (see section \ref{sec:nonanalyticity}).

For finite $N$, we consider this connection between topology and analyticity to be generic. It is a deeper question whether or not the nonanalyticity survives the thermodynamic limit. The topological approach is an attempt to answer this question and to establish a connection between the topology of the state space submanifolds $\Sigma_\varepsilon$ and thermodynamic quantities of the system. For systems like the mean-field spherical model, where nonequivalence of the statistical ensembles is observed, one would expect a correspondence between topological and thermodynamic quantities only with regard to {\em one}\/ of the statistical ensembles.

Comparing the topology of $\Sigma_\varepsilon$ in (\ref{Sigma_topo}) with the {\em canonical}\/ thermodynamics of the mean-field spherical model with zero external field $h=0$, one would associate the phase transition observed in (\ref{evonbeta_h=0}) at inverse temperature $\beta=1$ and internal energy $\bar{\varepsilon}=0$ with the topology change from $\S^{N-2}+\S^{N-2}$ to $\varnothing$ at $(\varepsilon,h)=(0,0)$. However, the same type of topology change is present for all $|h|<1$ at $(\varepsilon,h)=(\frac{h^2}{2},h)$, whereas canonical thermodynamics asserts the absence of a phase transition for all $h\neq0$. Hence, one would conclude that the information contained in the topology of $\Sigma_\varepsilon$ (or, in a related description, of $M_\varepsilon$) might not be sufficient as to distinguish between the presence or absence of a phase transition.

However, a meaningful connection between the topology of $\Sigma_\varepsilon$ and the thermodynamics of the model can be established in the {\em microcanonical}\/ ensemble. Within this framework, following the discussion of section \ref{sec:nonequivalence}, the zero field case is not special, as a phase transition occurs neither with nor without an external field. This is in accordance with the topological results where the topology change is the same for all $|h|<1$.\footnote{Note, however, that for a class of systems with {\em non-confining}\/ potentials the information from the submanifold topology was found to be insufficient as to distinguish between the occurrence and the absence of a phase transition, regardless of the statistical ensemble considered.\cite{Kastner}}

The mean-field spherical model is not the only example for which the observed topology changes find a meaningful interpretation only when compared to microcanonical quantities. For the mean-field $XY$-model as considered in references \cite{CaPeCo2} and \cite{CaCoPe}, the observed discontinuity in the modulus of the Euler characteristic is interpreted as being related to the phase transition of the model at zero external field $h=0$ in the canonical ensemble. However, similarly to the mean-field spherical model, the same topology change is present for $h\neq0$, although a phase transition does not occur in this case. Again, a meaningful interpretation of these findings should be possible when comparing topology to microcanonical thermodynamics.

In addition to the above observations based on examples, a general remark on the relation between the topological approach and the statistical ensembles can be made. The microcanonical ensemble and the topological approach, different as they are, have the same quantity as a starting point: The microcanonical partition function (\ref{Omega_general}) of a system, which is at the basis of the microcanonical ensemble, can be defined as the {\em volume}\/ of the level sets (\ref{Sigma_def}). Considering the {\em topology}\/ of the very same level sets, we arrive at the fundamental quantity underlying the topological approach, and, in the light of this observation, the close connection between these two concept appears plausible.

\section{Summary and conclusions}

The mean-field spherical model has proved to be an instructive and profitable laboratory for the investigation of a variety of effects and theories in the field of the statistical mechanics of phase transitions. Due to its simplicity, an exact solution can be obtained for any finite or infinite number $N$ of degrees of freedom, even in the presence of an external magnetic field.

An exact expression for the microcanonical entropy density $s$ as a function of the energy density $\varepsilon$ was obtained. For external magnetic field $h$ with $0<|h|<1$, $s$ is found to be nonanalytic at $\varepsilon=|h|-\frac{1}{2}$ for arbitrary finite $N$. This is in contrast to the canonical case, where the free energy density $f_N$ as a function of the inverse temperature $\beta$ is real-analytic for all finite $N$.

Also for the canonical partition function $Z_N$ as a function of $\beta$ an exact expression was given, allowing a discussion of the {\em loci}\/ of the zeros of the partition function in the complex $\beta$-plane. For $h=0$ and in the limit of large $N$, these zeros are found to lie dense on certain branches of the Szeg\"o curve, approaching the real axis at $\beta=1$. The observed pinching of the real axis is a beautiful illustration of the Yang-Lee-mechanism of analyticity breaking in the thermodynamic limit.

Equivalence of the mean-field spherical model of finite system size $N$ to a mixed isovector/isotensor $\sigma$-model on a lattice of two sites was established. This result implies equivalence of the mean-field spherical model with zero magnetic field to the $\RP^{N-1}$ $\sigma$-model.

By comparing, for $h=0$, the microcanonical and the canonical solutions of the mean-field spherical model in the thermodynamic limit $N\to\infty$, ensemble equivalence is found to be violated. Instead, a phenomenon termed {\em partial equivalence}\/ is observed. This can be viewed as a consequence of the fact that the microcanonical entropy $s$ is a strictly concave function with compact support, while its derivative, the microcanonical inverse temperature $\beta(\varepsilon)=\frac{\partial s(\varepsilon)}{\partial \varepsilon}$, has a positive lower bound.

A complete characterization of the topology of the state space submanifolds $\Sigma_\varepsilon$ of constant energy $\varepsilon$ is possible for the mean-field spherical model. The result allows for a discussion of the relation of these topological quantities to the thermodynamic behavior of the system. Based on this example, the connection between topological and thermodynamic quantities in the presence of ensemble nonequivalence was discussed, observing a close relationship between the topological approach and the microcanonical ensemble.

\section*{acknowledgments}
M.\ K.\ acknowledges financial support by the Deutsche Forschungsgemeinschaft (grant KA2272/2).

\end{document}